# Enhancement from plasmonic-molecular coupling for mass transduction

Giuseppina Simone

**Abstract**

The plasmon-mechanical resonators are frequently used in the development of sensors. Active frameworks impose mechanical motion into the lasing dynamics through the use of an optical gain and achieve better sensitivity. Here plasmon-mechanical coupling is demonstrated in a multilayer when a surface plasmon polariton/Fabry-Pérot hybrid mode is excited in a Kretschmann configuration, while it is observed that the strong plasmonic dispersion allows the deformation of the mechanical domain at several frequencies. After a dye is adsorbed on the surface of the cavity, the layout of the optomechanics is schematized by a spring-mass oscillator mounted onto the surface of the cavity-end mirror. The system is defined by its capability to determine the experimental settings with the best resolution before a controlled experiment in which the oscillator senses a mass. The advantages and disadvantages of the procedure are presented once the data have been assessed and modeled.

## I. INTRODUCTION

Numerous sensing devices based on the coupling of the optical and mechanical domains [1–4] [5,6] in the form of optomechanical resonators (such as cantilevers[7] and cavities [8]) have been developed during the past decade [1] as a result of recent micro and nanooptics advances. The optomechanical coupling is classified as dispersive and dissipative [9], depending on whether the mechanical resonator modulates the resonant frequency by changing the cavity length or the decay rates by altering the optical input coupling or intracavity loss and these variations measure the system sensitivity. Interferometers in the form of Fabry-Pérot cavity are optomechanical resonators [10] [11] that are frequently used as extremely accurate force detectors [12,13]. They are made to deform while under stress, with the sensitivity measured by the number of deformations [14,15], whereas high-frequency transduction can enable the sensing of mass changes or small forces [16] [17]. A change in the cavity length affects the optical transmission and reflectivity, which causes a modification in the applied force through the cavity length [18,19], while a change in the laser intensity and frequency causes a variation in the optomechanical coupling. A quantum mechanical ground state is attained through a mechanism of mechanical vibration cooling when the coupling between the mechanical and optical domains takes place in the resolved sideband limit defined by a hierarchy between the frequency of the mechanical oscillator and the cavity loss rate, $\omega_c \gg \kappa$; this condition is related to a strong optomechanical coupling and causes several relevant advantages in sensing. In reality, Fabry-Pérot cavities have modest mode volume and require a high mechanical mode quality factor and ultra-narrow frequency light sources for resulting in an ultrasensitive mass/displacement force transduction, potentially important in single-molecule research [20]. Hybrid cavities developed with the contribution of plasmonic antennae, e.g. fluorescent molecules, quantum dots, and photo-switchable molecules, enable a strong coupling by the concentration of the electromagnetic field to deep sub-wavelength achieved by exciting the surface plasmons that then results in small-volume modes and compensates the metallic loss contribution. To transfer mechanical oscillation onto lasing dynamics generated by an interaction between the gain particle-photon and photon-phonon, simplified architectures have been redesigned with a spring-mass oscillator mounted onto one end mirror of the cavity [21,22]. Shift, spectrum broadening, temporal coherence, and photon statistics depend on the optomechanical system in these systems [23]. Besides, a coherent and reversible energy transfer faster than the dissipation between the emitter and the cavity as well as the creation of two new hybrid eigenstates are all results of the strong coupling between the molecules and the hybrid system.

Here, experimental and numerical research involving the coupling of a passive and lasing cavity was carried out, where a multilayer consisting of two mirrors (indium tin oxide ITO and silver Ag) interacts with a molecular dye. Initially, an analysis aimed at examining the Fabry-Pérot hybrid/surface plasmon polariton SPP hybridization in the near-field region was developed; then, the dye (rhodamine G6 [21]) adsorbed on the Ag modifies the optomechanical spectrum and leads to a coupling rate greater than the optical decay ($\kappa^* < \omega_c$) [24]. The hallmark of the emitter's contribution to enhancing the sensitivity was characterized; hence, mass sensitivity was monitored by the frequency shift as various masses of deionized water were tested.

In conclusion, the findings point to a system that can measure both high and low-speed phenomena by coupling a high-frequency mode of the order of a few terahertz from the molecular emitter with a low-frequency mode of the order of a few megahertz from the cavity [25,26] and prospect an impact on molecular sensing, which still exhibits inefficiencies [27,28].

## II. EXPERIMENTAL

ITO-coated cover slides were utilized as the substrate, and the (3-mercaptopropyl) trimethoxysilane (shortened as //) (Sigma Aldrich, China) was grown by chemical vapor deposition in a furnace including a vacuum, temperature, and pressure control (APPENDIX I) [29]. Table 1 shows some properties of the multilayer. For the measurements, the laser (Shenzen Futhe Tech. Co. Ltd, 5mW, aperture 1 mm) was collimated (Thorlabs RC12FC-P01) and polarized by a double Glan-Taylor Calcite Polarizer; the noise was lowered with an optical iris diaphragm (Thorlabs, D25SZ). Before being collected by a silicon Si photodiode and delivered to an oscilloscope, the reflected beam was collimated and filtered (Thorlabs, Glan Thomson polarizer, 650-1050 nm). For an optomechanical analysis, the front portion of the prism (Thorlabs; UV Fused Silica 25 mm right) held a Si photodiode ($\lambda_0$ = 960 nm bandwidth) and a ceramic disk capacitor for noise control [30–32].

**Table 1.** Layout and size of the multilayer

| Layer | Material | Thickness μm | Mass Density kg m$^{-3}$ | Poisson Ratio | Young Modulus GPa |
|---|---|---|---|---|---|
| 1 | ITO | 0.080-0.1 | $1.18 \times 10^{11}$ | 0.2 | 7140 |
| 2 | SiO$_2$ | 500 μm | $7 \times 10^{10}$ | 0.23 | 2500 |
| 3 | (3 mercaptopropyl) trimethoxysilane | $1 \times 10^{-6}$ | | | |
| 4 | Ag | 0.050-0.070 | $8.3 \times 10^{10}$ | 0.37 | 10490 |

Matlab was used to create the numerical analyses (Ed. 2023). The system's finite element analysis was modeled with the help of the partial differential toolbox. Under specified boundary conditions and by diagonalizing the coupled linear equations of motion, the modal analysis provided the eigenfrequencies and the corresponding stress and strain fields of the meshed structure. The deformation that results from the model's plasmonic and mechanical interaction was constructed assuming stress applied to the inside of the two mirrors equal to $\sigma = 400\ MPa$. Fused silica, ITO, SiO$_2$, and Ag made up the model's layers, with silver described as a homogeneous layer while neglecting its actual properties (APPENDIX I Figure S1). The refractive index was acquired from Palik [33] (Ag, ITO) and Schott (fused silica) in addition to the values listed in Table 1, and experimental parameters were utilized to calibrate the model [34]. The transfer matrix method was used to model the dispersion when a *p*-polarized light incident from the prism. The incident plane wave complex amplitude of the electric field phasor was linked to the complex amplitudes of the electric field phasor of the reflected and transmitted plane waves by $4 \times 4$ matrix per layer.

**BACKGROUND AND DESIGN**

ITO/SiO$_2$//Ag Fabry-Pérot cavity was mounted onto a prism to measure the reflectance in an angular space described by $\vartheta$ (Figure 1a). The laser beam travels through the prism, the transparent ITO and SiO$_2$, before generating the surface plasmon polaritons at the Ag layer (Figure 1b). The mechanical oscillator supports hundreds of modes (Figure 1c) with frequencies upwards of several hundred hertz and the first one at $\Omega_m = 600\ Hz$ with a quality factor $Q_m = \frac{\Omega_m}{\Gamma_m} = 30$ ($\Gamma_m = 20\ Hz$) [29]. The optomechanical coupling factor for a Fabry-Pérot cavity is $G = -\omega_c/L$ [35]; photons inside the cavity exert a force on the inner face of the mirrors resulting in a radiation pressure ($F_{rad}(t) = -\hbar\ G$) responsible for a transfer from light momentum to the mechanical system inside the optical cavity and for an optical-mechanical parametric coupling (Figure 1d). Optomechanics depends on the activation of an optical cavity by a laser beam whose central frequency is detuned from the cavity mode's resonance frequency [36] [37]. The nonzero mechanical displacement $x(t) \neq 0$ is accounted as a measurement of the sensitivity because it perturbs the electromagnetic cavity and shifts the optical resonance of a quantity $\Delta\omega_c = \omega_c(t) - \omega(0) = Gx(t)$. The displacement is described by a three-dimensional vector field that drops to one-dimensional domain by mapping the vector field to a scalar displacement $x$, which is associated to the mode energy $U = \frac{1}{2}m_{eff}\Omega_m^2 x^2$ through the effective mass $m_{eff}$ and frequency $\Omega_m$ of the oscillator. A more stable figure of merit for determining the strength of the coupling between the optical and mechanical modes results in a vacuum optomechanical coupling rate $g_0 = G\ x_{zpf}$ [38] that by means of the mechanical oscillator zero point fluctuations $x_{zpf} = \sqrt{\hbar/2m_{eff}\Omega_m}$ overcomes the ambiguity related to the determination of the parameter $G$, bypasses the shift $\Delta\omega_c$ of the cavity frequency and the measurement of the displacement of the structure at a given position.

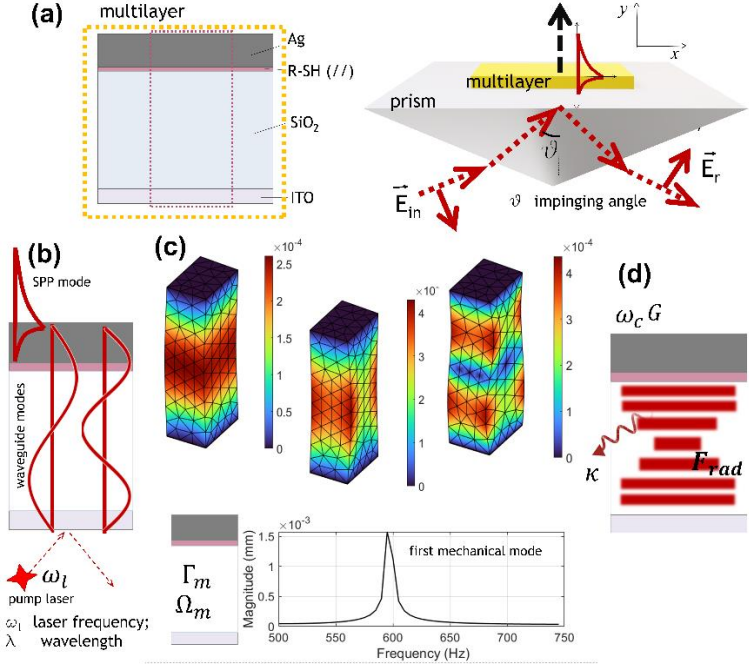

**Figure 1.** Plasmon-mechanical coupling. (a) Multilayer layout; right: Kretschmann configuration. (b)Mul Modes of the multilayer. FP: Fabry-Pérot. (c) Top: Three eigenmodes. Bottom: first mode amplitude and scheme. (d) Optomechanics.

### III. RESULTS

**A. Characterizing the plasmonic behavior**

For a few different $SiO_2$ layer heights, ITO/$SiO_2$//Ag dispersion as a function of the impinging angle was studied and the results were contrasted with the dispersion of $SiO_2$//Ag design (Figure 2a). A surface plasmon polariton SPP mode characterizes both the layouts of $SiO_2$//Ag (panel (i)) and ITO/$SiO_2$//Ag (panels (ii-iv)); though, when the ITO layer was added to the system's architecture, the perfect metal/lossless dielectric/absorptive ITO arrangement generates waveguide modes (square in panels (ii-iv)) [39,40] [41] [42] with an aptitude to hybridize the SPP mode. The hybridization results are more clearly characterized at 500 $\mu m$ -$SiO_2$ thickness (panel (iii)) than they are at other dielectric heights (panels (ii), (iv)); in turn, the plasmon wavevector (dashed line, panel (i)) is height-invariant [43]. The experimental reflectivity displays that $SiO_2$//Ag's angular reflectance spectrum showed a single main mode at $\vartheta = 40\ deg$ 40, while the multilayer's reflectivity spectrum showed two waveguide modes at $\vartheta = 44.9\ deg$ and $\vartheta = 48.6\ deg$, as well as a weaker peak at $\vartheta = 40\ deg$ (Figure 2b). Kretschmann geometry enables the mode tuning and hybridization [44] pivotal for a strong exciton/polariton coupling [45]. The mode at low angular aperture depends on the Ag layer, while the high angular aperture waveguide mode forms that undergo hybridization. At $\vartheta = 44.9\ deg$, the enhancement $\frac{|E_z|^2}{|E|^2}$ (panel (b) right) shows the waveguide-SPP mode hybridization. Hence, the cavity operates as a photodiode; the built-in potential difference is provided by ITO/$SiO_2$, while $SiO_2$ acts as a cavity having a thickness enabling the optical path length to support a waveguide mode and its hybridization with the SPP mode. Figure 2c illustrates the analysis of the reflectance in the wavelength framework according to the impinging angle, and Figure 2d exemplifies the calculated resonance frequency of the Fabry-Pérot mode as a function of cavity width and the differential peak shift (right y-axis) at each cavity width. Finally, by combining the effective volume $V_{eff}$ from geometrical considerations ($V_{eff} = 5 \times 10^{-16} \, m^3$) and the diffraction volume $V_0 = 3\frac{(c/f)^3}{4\pi^2}$ ($= 7.14 \times 10^{-21} \, m^3$; $f$ is approximated by frequency @ 633nm in air, while $c$ is the speed of light in SiO$_2$, $c = 2 \times 10^8 \, m\, s^{-1}$), the volume mode factor $V = \frac{V_{eff}}{V_0}$ was estimated $V = 7 \times 10^4$.

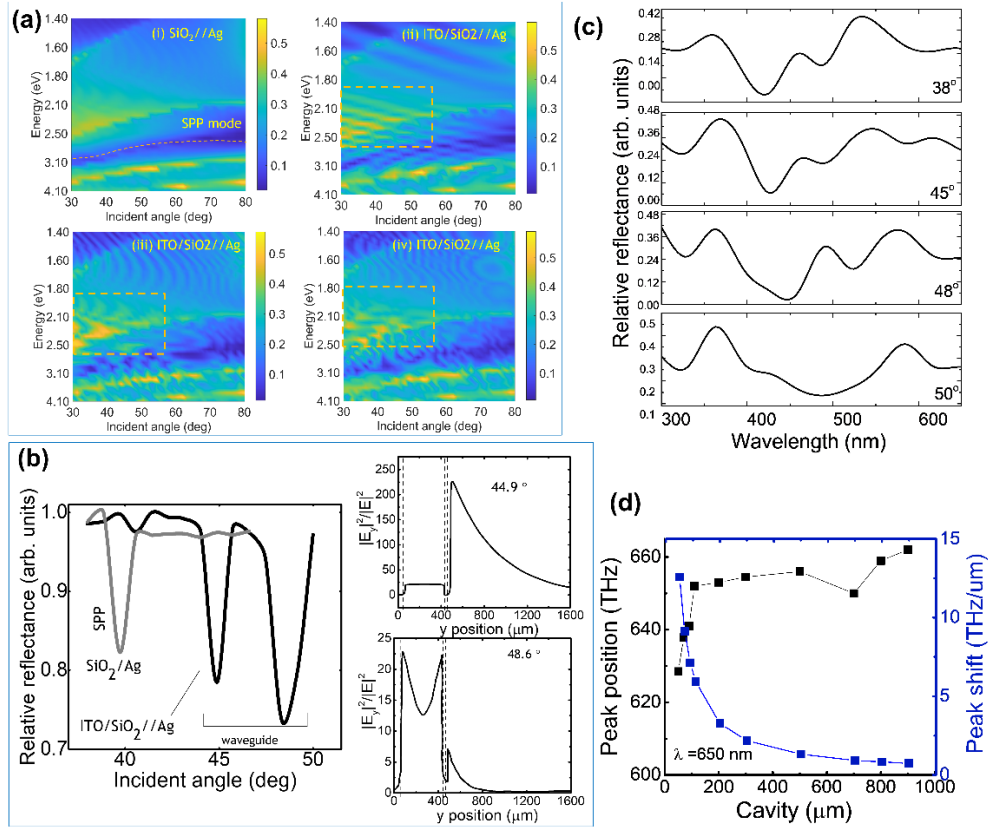

**Figure 2**. Plasmonic characterization. (a) Reflectivity dispersion. SiO$_2$ thickness (ii-iv): clockwise 100 μm, 500 μm, and 700 μm. Color bar: reflectance intensity. (b) Experimental relative reflectance. Right: *y*-profile of the electric field enhancement $\frac{|E_z|^2}{|E|^2}$ of ITO/SiO$_2$//Ag layout for the modes at $\vartheta = 44.9 \, deg$ and $\vartheta = 48.6 \, deg$. (c) Relative reflectance of ITO/SiO$_2$//Ag layout (Table 1). (d) Calculated resonance frequency as a function of cavity width of the Fabry-Pérot mode at $\vartheta = 44.9 \, deg$ (left *y*-axis) and differential peak shift (right *y*-axis) at each cavity width.

### B. Plasmon-mechanical coupling

The optomechanical mode was turned on using the optical power of a 5-mW narrowband laser (short-term line width <100 kHz, $\lambda = 633 \, nm$, TM polarization) aimed into the center of the on-resonance multilayer. A spectrum analyzer converts into the frequency domain the transmitted signal collected by the photodiode and sent to an oscilloscope (Figure 3a). The spectra at various impinging angles, including the angle of resonance, demonstrate the mechanical mode invariance according to frequency and the wavevector (Figure 3b). Peak splitting and increased intensity at $\vartheta = 44.9 \, deg$ are highlighted by focusing the inspection on certain frequency ranges (panels 1, 2, and 3); in addition, power-dependent measurements of the resonant spectrum evidence the main mechanical mode growth for raising DC power (Figure 3c). By aiming the laser beam at the photodiode at an angle of $\vartheta = 44.9 \, deg$, the laser contribution was read; the signal (grey curve, Figure 3d) has a signal-to-noise ratio of $40\ dB$ and varies as $1/f$ at low frequencies before collapsing to a constant value at higher frequencies. Recent studies have demonstrated that when a cavity-emitted signal is delivered through narrowband optical filters centered on the resonance, the pump is reduced and photon counting events directly correspond to counting phonon absorption events (for example, emission) [46,47]. Because the peak area estimates the mechanical mode phonon count, through a photon/phonon swap, the mechanical cavity interior photon count increases as the system approaches the plasmonic resonance, according to the results.

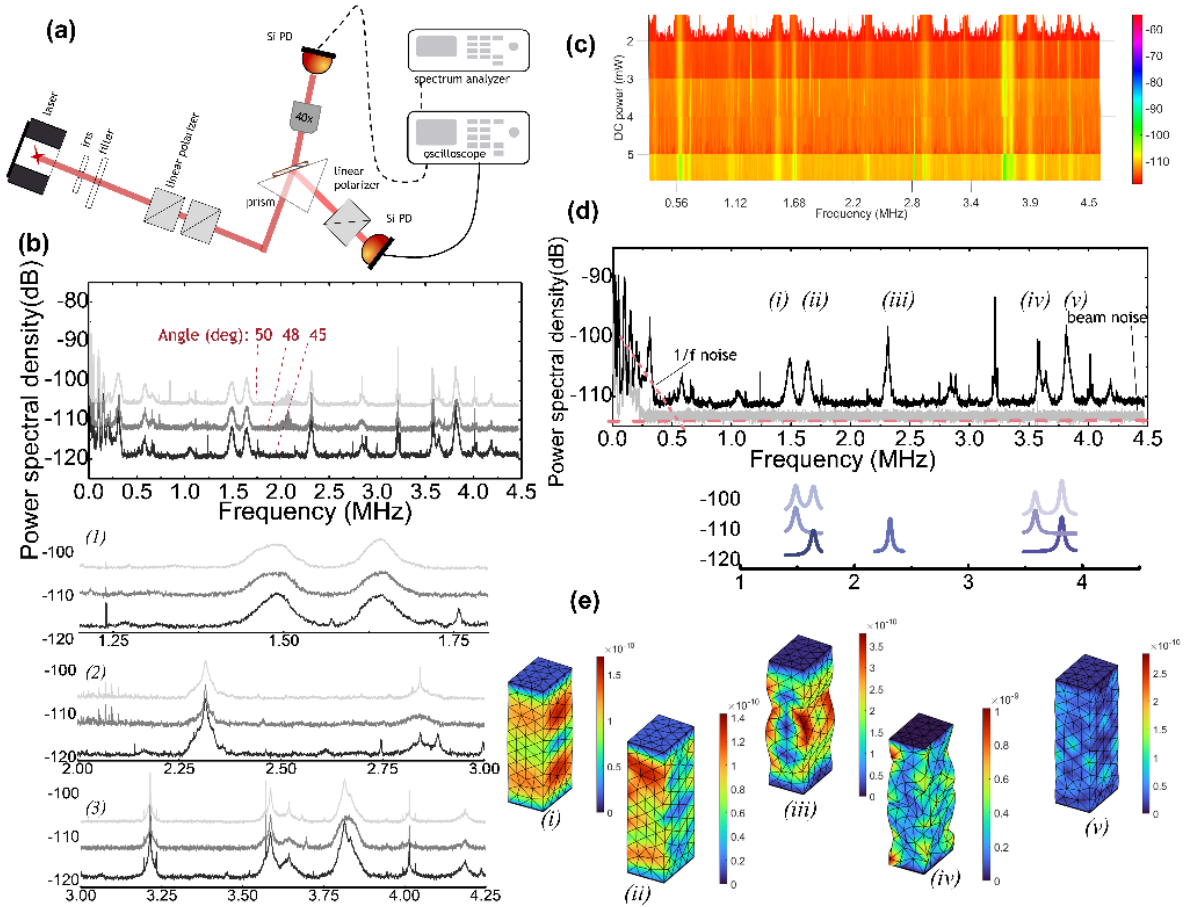

**Figure 3**. Plasmon-mechanical coupling. (a) Experimental setup. Laser ($\lambda = 633\ nm$) input power and polarization are controlled by using a halfwave plate followed by a linear polarizer. The reflected and transmitted light is sent to fiber-coupled amplified photodiodes. The AC output is monitored on a spectrum analyzer, while the DC output is monitored on an oscilloscope. (b) Power spectral density of the multilayer at different impinging angles; bottom: frequency sub-ranges. (c) Power-dependent mechanical spectra. (d) Spectrum at $\vartheta = 44.9\ deg$; the main modes are labeled. Bottom: Lorentzian fitting. (e) Eigenmodes from finite element model simulations.

A single mode (iii), located in the middle of the frequency range, and side modes ((i)-(ii);(iv)-(v) Figure 3d) are traits associated with the multilayer trace. These side modes are distinguished by a similarity of the quality factor (Table 2), which highlights the split resonance doublet trait and is attributed to a degeneracy excitement of the resonant modes caused by a dispersive coupling, in contraposition to dissipative coupling played by the resonant mode (iii) [9] [18]. By applying stress to the inner section of the two mirrors, a finite element analysis produces coherently, in-plane eigenmodes at the doublets and a dissipative out-of-plane deformation at 2.3 MHz (Figure 3e).

**Table 2.** Properties of the mode estimated by Lorentzian model

| Label | $\omega_c$, MHz | $\Gamma$, MHz | Q | U, MHz |
|---|---|---|---|---|
| (i) | 1.49 | 0.060 | 24.9 | 0.42 |
| (ii) | 1.64 | 0.059 | 27.9 | 0.38 |
| (iii) | 2.31 | 0.036 | 63.9 | 0.52 |
| (iv) | 3.59 | 0.048 | 74.9 | 0.40 |
| (v) | 3.82 | 0.049 | 77.5 | 0.63 |

The frequency shift of the plasmonic peak is $2.5 \frac{THz}{\mu m}$ (Figure 2d) with the energy $U$ stored in each mechanical mode achieved by Lorentzian fits (Table 2). Through $U$, the effective mass can be computed (e.g. (ii): $m_{eff,ii} = 11.9\ pg$). At $\vartheta = 44.9\ degrees$, the data shown in Figure 2c provide an estimation of the frequency shift $G = 1.3 \times 10^{18}\ Hz/m$ and of the radiation pressure $\hbar G = 2\ GHz/nm$; it follows that the root-mean-square amplitude of the mechanical oscillator zero point fluctuations results in $x_{zpf} = 0.0013\ nm$ with a coupling rate $g_0 = 0.17\ MHz$. The highest coupling factor is visible at $3.59\ MHz$ when the study is expanded to include all peaks ((i)-(v)).

### C. Characterizing the sensitivity of the sensor

Rhodamine G6, a fluorescent dye that may make an excellent molecular emitter, was adsorbed as a radiative antenna onto the surface of the Ag layer. Because of the aggregated micrometric features that make up the Ag layer in the real application (Figure 4a), when the dye was applied to the surface, it soaked through the gaps between the Ag and SiO$_2$ and dried. The upgraded optomechanics is based on a molecular spring-mass oscillator that connects to one end of the cavity mirror (Figure 4a, right). A pump laser causes the molecules to transition from a ground $|g\rangle$ to an excited $|e\rangle$ (Figure 4a, bottom). At an emitting frequency $\omega_m$, the dipole shows an intrinsic radiative decay rate $\gamma_m$ influencing the behavior of the simple optomechanical cavity and a negligible non-radiative decay rate. The cavity in turn has a plasmonic loss $\gamma_{spp}$ at frequency $\omega_{spp}$ and, because of the crystalline Ag, loses a part of its energy to heat at a rate $\gamma_{ohm}$. Through the molecular dipole and geometrical factors, the decay rate modifies as $\kappa^* = \frac{V_{eff}}{V_0}\frac{\gamma_m^2}{\gamma_{spp}}$ [48], which in the nearfield assumes a value $\kappa^* = 1.9\ ps^{-1}$ corresponding to a quality factor $Q = \frac{\omega_{spp}}{\kappa^*}(= 1.1 \times 10^9)$ (Table 3). According to the equation $\gamma'_m = \gamma_m\ Q/V$, the cavity mode volume and the quality factor modify the radiative rate of the molecular dipole. Besides, in the updated optomechanics, the cavity/dye has a nearfield radiation rate schematized as $\gamma_f = \gamma'_m \frac{\kappa^*}{\kappa^* + \gamma_{ohm}}$ and, altogether, one can conclude that radiative rate outperforms the cavity decay of a factor $\frac{\gamma_f}{\kappa^*} = 2.3$.

Intrinsic mechanical dissipation, optical noise, and molecular damping contribute to the noise of the optomechanics. Spreading below the combined plasmon-mechanical noise from the intrinsic mechanical dissipation, the shot noise power spectrum is negligible (Figure 3c); in turn, the molecular damping and mechanical dissipation are combined in Figure 4b. The molecular damping-to-mechanical dissipation ratio can be considered equivalent to the damping rate ratio, according to the dissipation-fluctuation theorem (Appendix II).

**Table 3**. Factors and values of the system

| Variable | [49] $\gamma_m$ | $\omega_m$ | $V$ | $\gamma_f$ | $\gamma_{ohm}$ | $\gamma_{spp}$ | $\kappa *$ | $Q$ | $\gamma'_m$ | $\omega_{spp}$ |
|---|---|---|---|---|---|---|---|---|---|---|
| Value | $285\ ps^{-1}$ | $5.5 \times 10^8\ MHz$ | $7 \times 10^4$ | $4.5\ ps^{-1}$ | $2 \times 10^6 ps^{-1}$ | $3 \times 10^9\ ps^{-1}$ | $1.9\ ps^{-1}$ | $1.1 \times 10^9$ | $4.7 \times 10^6 ps^{-1}$ | $2.2 \times 10^9\ MHz$ |

Therefore, the intrinsic mechanical dissipation and molecular damping were separated (Appendix II Figure 2), and molecular damping of 68 kHz and an intrinsic mechanical dissipation of 12 kHz were measured. The molecular damping dominates by a factor of 5.6 and the equivalent noise of the sensor is about 15% of the noise derived from the vibration of the molecules and interaction with the sensing element.

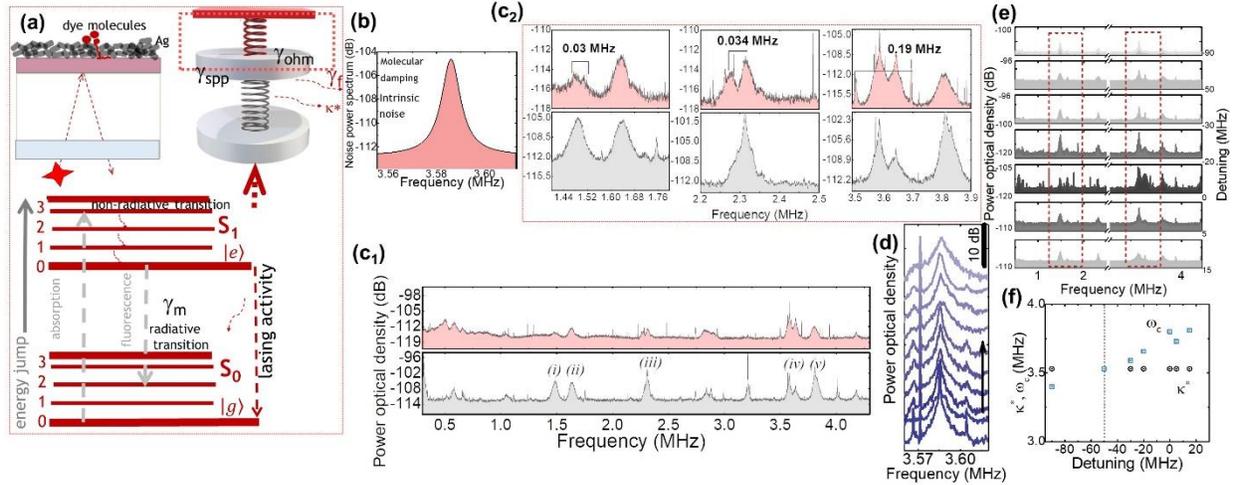

**Figure 4**. Plasmon-mechanical coupling with a dye-oscillator. (a) Left top: Schematic of the experimental setup; right top: external oscillator connected to the cavity (out-of-scale and proportions); bottom: lasing activity of the dye. (b) Noise spectral density of the complex system including the intrinsic damping. (c$_1$) Power density of before (bottom) and after (top) the dye adsorption; (c$_2$) doubling of the mode peaks (d) Power optical density of (iv) according to the pump power (from bottom to top: from bottom to top: 0.5-5 mW, Δ=0.5 mW. (e) Power optical density according to the laser detuning. (f) Comparison of the optical ($\kappa^*$) radiation rate with the coupling rate ($\omega_c$) as a function of detuning.

The spectrum of the multilayer with the dye (Figure 4c$_1$) displays split resonance doublets at (i),(iii) and (iv) (Figure 4c$_2$), which derive from an exciton/polariton strong coupling (APPENDIX III) [50]. The $g_0$-highest mode-(iv)-displays the highest splitting; in addition, by raising the pump power, the peak grows concomitantly with a frequency merging of the three eigenmodes (Figure 4d). Figure 4e depicts the mode detuning series following a splitting frequency trend and Figure 4f compares the plasmonic decoherence rate with the coherent coupling rate as a function of the detuning and shows the hierarchy $\omega_c > \kappa^*$ above the resonance caused by the absorbed light large amount. The cavity reaction to dye concentrations as a function of the force applied by the molecule on the cavity and the influence on the electromagnetic field also serves to illustrate the plasmon-mechanical sensitivity (Appendix IV).

When the mass sensor responsivity ($\mathcal{F} = \frac{\partial \omega_c}{\partial m_{eff}}$) was evaluated (plain cavity, mode *(ii)*, $\mathcal{F} = 10^{17}\ Hz\ g^{-1}$), deionized water was used as an example due to *in-vivo* biology assessment relevance of mass responsiveness in an aqueous environment. Figure 5a$_1$ depicts the mechanical resonator frequency shift as a function of sample volume, while Figure 5a$_2$ focuses on the mode at 0.5 MHz frequency shift and mode (iv). The shift was caused by the mechanical resonator's increasing mass, and there is a direct linear relationship between the frequency shift and the mass acting on the mechanical resonator (Figure 5b). As demonstrated by a comparison with an identical measurement taken from a plain cavity (Figure 5c), the dye amplifies the measured change by at least 25%.

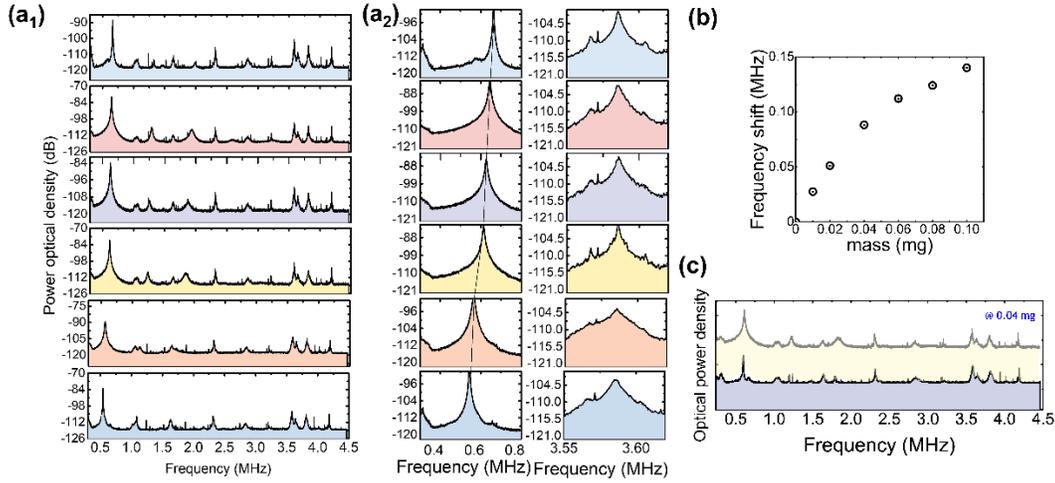

**Figure 5**. Mass sensitivity. (a$_1$) Optical density for different water masses; from bottom to top: 0.01 mg, 0.02 mg, 0.04 mg, 0.06 mg, 0.08 mg, 0.1 mg; samples were applied on the mechanical resonator with a high-precision pipette. (a$_2$): peak at 0.5 MHz, and peak (iv). (b) Shift/mass relationship, mode at 0.5 MHz. (c) Comparison between the plain cavity and dye-cavity; water mass: 0.04 mg.

## V. CONCLUSIONS

An optomechanical oscillator was examined where the SPP/waveguide Fabry-Pérot mode was coupled to a mechanical mode. Experimental and numerical findings established that the strong plasmonic dispersion allows the deformation of the mechanical domain at several frequencies. A spring-mass oscillator mounted onto one end mirror schematizes the optomechanical arrangement following the adsorption of a dye on the Ag surface. The experiments on the actual device showed that the strategy increased the radiation rate by a factor $\frac{\gamma_f}{\kappa^*} = 2.3$ and impacted the system performance. Through the frequency shift, the sensor mass capability was measured.

The active optomechanical design here depicted takes advantage of two separate frequency ranges: the ultra-high one in the range of terahertz typical of the dye and the low one in the order of megahertz typical of the cavity. Therefore, the resonator enables high-speed operation that is faster than is currently possible with optomechanical sensors [25,51] and enables the tracking of physical interactions occurring over brief time intervals; moreover, the high-frequency resonator with a low Brownian motion enables the measurement of low-frequency vibration modes caused by dissipative forces. Because they provide information about conformational changes and their functional significance as well as useful information about a variety of disorders, these forces are vital to molecular physics [52]. Identifying low-frequency mechanisms require high sample concentrations or optical oscillators with high sensitivity and the capacity to detect biomolecular changes because of the electric transition signal dominance over the excitation coefficient in biological events. Force nanoscale sensors may be able to record physical interactions that take place over incredibly tiny intervals at rates that are substantially faster, however, they frequently require a lot of resources. The accuracy of optical approaches combined with mechanical devices that are ultimately quick and perturbation-free may significantly broaden the spectrum of applications.

# Supplementary Material

Experimental and numerical results to complement and support the results are provided

| Symbols | Definition |
|---|---|
| $\Omega_m$ | mechanical frequency |
| $\Gamma_m$ | mechanical damping rate |
| $Q_m$ | mechanical quality factor |
| $\omega_{spp}$ | plasmonic frequency |
| $\gamma_{spp}$ | plasmonic loss |
| $\omega_c$ | optomechanical frequency |
| $\omega_m$ | molecular frequency |
| $G$ | optical frequency shift |
| $g_0$ | optomechanical coupling rate |
| $x$ | mechanical displacement |
| $F_{rad}$ | radiation pressure |
| $x_{zpf}$ | mechanical zero-point fluctuation amplitude |
| $m_{eff}$ | effective mass |
| $U$ | mechanical mode energy |
| $V = \frac{V_{eff}}{V_0}$ | mode volume rate |
| $V_0$ | diffraction volume |
| $V_{eff}$ | effective volume |
| $\kappa$ | optomechanical radiation rate |
| $\kappa^*$ | modified optomechanical radiation rate |
| $\gamma_f$ | nearfield radiation rate |
| $\gamma'_m$ | modified molecular radiation |
| $\gamma_m$ | molecular radiation |
| $\gamma_{ohm}$ | Ohmic loss |
| $\vartheta$ | impinging angle |
| $\lambda$ | pump excitation wavelegth |
| $\omega_l$ | pump excitation frequency |

APPENDIX I

Ag features have been produced using a two-stage process, with the first phase based on the generation of particles as clusters of nucleation and the second on the growth of the particles.

The chemicals were used without any purification. NaCl (>99%), AgNO3, polyvinylpyrrolidone (PVP, 40000 MW), and ethylene glycol (anhydrous, 99.8%) were purchased from Sigma-Aldrich. Ultrapure water with a resistivity of 18.2 MΩ cm was obtained from an mQ Integral Water Purification System from Merck Millipore.

To prevent effects due to the photosensitivity of the AgCl, the synthesis of the salt was performed in a dark environment. A silver nitrate aqueous solution (1 mL, 0.5 M) was mixed with a sodium chloride aqueous solution (1 mL, 1 M). (van de Donk et al. 2019; Zhang and Simone 2019). Following the addition of sodium chloride, flocculation started. The precipitate was separated by the supernatant and washed with fresh and ultrapure water before it was dried. The preparation of the Ag multishape features was performed in a three-neck round-bottom flask at 160 °C and under stirring. The PVP (0.2 mg) was dissolved in 10 mL of ethylene glycol and mixed until reaching a stable temperature; then, the AgCl was added all at once. Inside the sealed flask, the synthesis was left to run for 30 minutes (Simone 2023). Figure S1 shows the details of the Ag particles. To prepare the plasmonic samples, the functional mode (3- mercaptopropyl) trimethoxysilane, which served as the functional molecule, was used to create a chemical bridge between the Ag features and the surface of an ITO/SiO$_2$ substrate. The (3-mercaptopropyl) trimethoxysilane (Sigma-Aldrich) was used to graft the Ag multishape features to create the optical resonators. It was deposited by chemical vapor deposition (5 μL at 70 °C and 14.7 psia for 30 min) in a homemade furnace that included a vacuum system and a temperature and pressure control. Cover slides with an indium tin oxide ITO-coated side were utilized as the substrate, and the (3-mercaptopropyl) trimethoxysilane was deposited on the SiO2. The substrates were cleaned with a mild stream of toluene or methanol and then dried in a gentle flow of air. The pure Ag feature solution was poured on top of the thiol and allowed to dry for 24 h at room temperature. The substrate was treated with ultrasonication for 30 min in water to remove the multishape features that were not grafted onto the substrate.

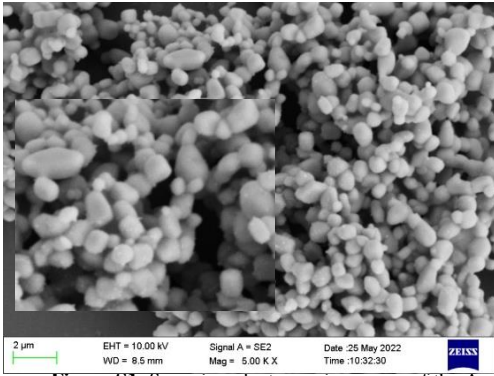

**Figure S1.** Scanning electron microscope of the Ag features.

APPENDIX II

The general equation of motion for harmonic oscillation $F_{ext} = m_{eff}\ddot{x} + m_{eff}\gamma\dot{x} + kx$, where $k$ is the spring constant and $\gamma$ is the damping rate and $m_{eff}$ is the effective mass. Three forms of damping describe the interaction of the multilayer with the dye (exciton/polariton coupling): intrinsic mechanical dissipation, optical noise, and molecular damping. The molecular and intrinsic mechanical damping introduce a combined force that opposes the velocity of the resonator. The plain multilayer has susceptibility $\chi(\Omega_m) = \frac{1}{m(\Omega_m^2 - \Omega^2 - i\Omega\Gamma_m)}$, where note that $\Omega_m$ is the natural frequency of the multilayer and $\Omega$ is the generic frequency, $m$ is the mass of the oscillator and $\Gamma_m$ is the damping rate. Under the action of a laser pump, the multilayer forms a coupling with the environment and transduces an externally applied force into a displacement. In a linear regime, the transduction is linear in the Fourier space: $x(\Omega) = \chi(\Omega)(F(\omega_c) + F_T(\Omega_m))$ where $x(\Omega)$ is the Fourier component of the multilayer displacement and $\chi(\Omega)$ is the Fourier component of the multilayer susceptibility, $F(\omega_c)$ and $F_T(\Omega_m)$ are the external force and the Longeving force resulting from the coupling of the multilayer and the pump laser. Then, when no external force is not applied, the Brownian motion induced by the Langevin force can be predicted according to the fluctuation-dissipation theorem $S_F(\Omega_m) = \frac{2Tk_B}{\Omega_m} Im\left(-\frac{1}{\chi(\Omega_m)}\right)$ and a thermal noise displacement $S_x(\Omega_m) = S_F(\Omega_m)|\chi(\Omega_m)|^2$; because of the high $\chi(\Omega_m)$, $S_F(\Omega_m)$ is negligible. The laser applies a pression into the cavity $S_{F_T}(\Omega_m)$, which causes a contribution to the damping of the emission of radiation rate. This strengthens the force noise density, with a damping $\kappa = \Gamma_m + \gamma_{spp}$. In addition, the molecular damping rate $\gamma'_m$ contribute to the displacement; and here, the equivalent damping is $\Gamma = \frac{\gamma'_m \gamma_{spp}}{\gamma'_m + \gamma_{spp}}$, neglecting the natural damping $\Gamma_m$. The damping rate of the molecules interacting with the multilayer is $\gamma'_m = \gamma_m Q/V$ (Kim, 2012) whereas $\gamma_m = l\frac{\omega^3 n|\mu|^2}{3\hbar\pi\varepsilon_0 c^3}$ where $\omega$ is the emission frequency, $n$ is the index of refraction, $\mu$ is the transition dipole moment, $\varepsilon_0$ is the vacuum permittivity, $\hbar$ is the reduced Planck constant, $c$ is the vacuum speed of light. Given that the thermal fluctuations introduced by the interaction with the molecules are independent of those introduced by thermal vibrations of the substrate and any other damping mechanisms intrinsic to the resonator, the total thermal force noise-power spectral density experienced by the resonator is $S_F(\Omega) =$ $2m_{eff}\Gamma T k_B = 2m'_{eff}\frac{\gamma'_m \gamma_{spp}}{\gamma'_m + \gamma_{spp}} T k_B$ and a thermal noise displacement $S_x(\Omega_m) = S_F(\Omega_m)|\chi(\Omega_m)|^2$. However, by considering the single two contributions of the damping, it is possible to assume $S_{Fspp}(\Omega) = 2m_{eff}\Gamma T k_B = 2m_{eff}\gamma_{spp} T k_B$ and $S_{Fm}(\Omega) = 2m'_{eff}\Gamma T k_B = 2m'_{eff}\gamma'_m T k_B = 2m'_{eff}\gamma_m l \frac{\mu}{m} T k_B$, so that

$$\frac{S_{Fspp}(\Omega)}{S_{Fm}(\Omega)} = \frac{\gamma_{spp}}{\gamma_m}\frac{m_{eff}}{m'_{eff} Q/V} \approx \frac{\gamma_{spp}}{\gamma_m}\frac{V^2}{\omega_{spp}}$$

where $C_m \sim 1$ is the ratio of effective masses before and after the dye adsorption, $\omega_{spp}$ is the SPP frequency and $V$ the mode volume as introduced in the main text.

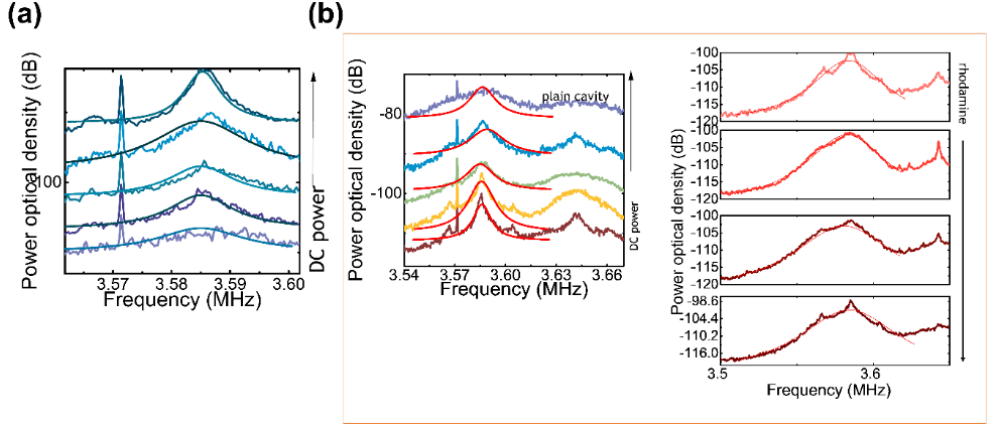

**Figure S2.** (a) Power spectra density versus frequency for several DC pumping powers (from bottom to top: 1, 2, 4, 5, 7 mW) of the plain multilayer. (b) Left: Power spectra density versus frequency for several DC pumping powers (from bottom to top: 1, 2, 4, 5 mW and plain cavity (top)) of the naïve multilayer including the dye. Right: Power density at different rhodamine concentrations, from top to bottom: 0.2 μM, 0.3 μM, 0.5 μM, 1 μM.

To determine the contributions to the noise from intrinsic mechanical dissipation and fundamental molecular damping, the following measurements were done. First, several DC pumping powers were applied to the plain cavity (Figure S2a). The damping rate of the resonance observed at 3.58 MHz was monitored. The decay rate plateaus to a minimum of 12 kHz, corresponding to the intrinsic mechanical dissipation $\Gamma_m$. For monitoring the rhodamine, different concentrations of the dye and different pump lasers like the one applied to the plain cavity were used (Figure S2b). With the rhodamine, the decay rate plateaus at 80 kHz. The difference between these two values corresponds to the molecular damping $\gamma_f = 68\ kHz$.

APPENDIX III

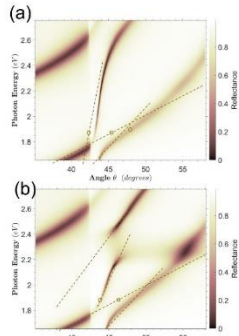

**FigureS3.** Numerical dispersion of the reflection according to the incident angle (a) plain multilayer and (b) with rhodamine. The numerical analysis simulates a TM domain as a function of the incident angle $\vartheta$, while keeping fixed the height of the layers

APPENDIX IV

The spectra's relationship to modes iv and v was investigated (Figure S4a). The stored optical energy and plasmonic mode resonance all changed when the concentration was increased (Figure S4b). According to the calibration curve in Figure S4b, the energy increases almost linearly at low concentrations, while doubling the concentration results in a plateau trend. The limit of detection, defined as the concentration given by the intersection of the calibration curve and the stability baseline, is determined by the relationship between the slope of the spectra and the concentration on the logarithmic scale (bottom Figure S4b). While the stability baseline (dashed line in Figure S4b, bottom) is defined as three times the standard deviation of the measurements of the sensor of a blank measurement. In accordance with the results, the limit of detection was found to be $10^{-4}$ mM.

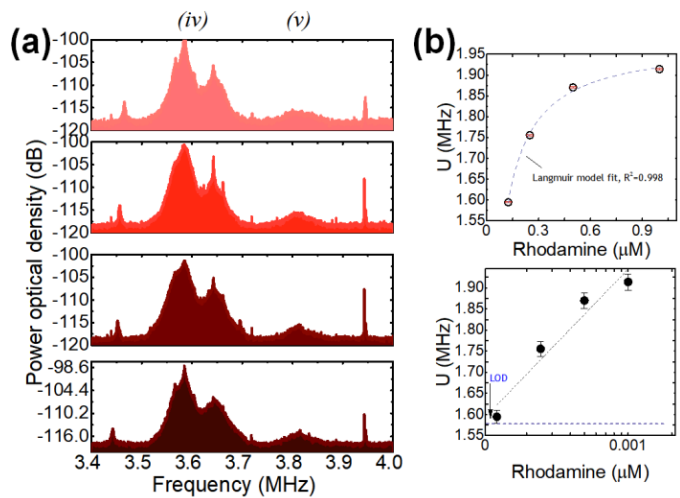

**FigureS4.** (a) Optical density for different concentrations of rhodamine with reference to the modes *(iv)* and *(v)*; concentration from top to bottom: 0.2 μM, 0.3 μM, 0.5 μM, 1 μM. (b) Top: linear relation between the mode energy U and the concentration (mode (iv)); bottom: limit of detection LOD estimation (mode (iv)).

**Publication bibliography**